\documentclass[10pt,conference]{IEEEtran}

\usepackage{cite}
\usepackage{amsmath,amssymb,amsfonts}
\usepackage{algorithmic}
\usepackage{graphicx}
\usepackage{textcomp}
\usepackage{xcolor}

\usepackage{booktabs}     
\usepackage{colortbl}    
\usepackage{xcolor}      
\usepackage{boldline}  
\usepackage{balance}
\usepackage{enumitem} 
\usepackage{fancyhdr}

\usepackage{wrapfig}  
\usepackage{siunitx}
\newcommand{\hpcacameraready}

\newcommand{\hpcayear}{2027}

\def\BibTeX{{\rm B\kern-.05em{\sc i\kern-.025em b}\kern-.08em
    T\kern-.1667em\lower.7ex\hbox{E}\kern-.125emX}}
\begin{document}
	
\definecolor{tablegray}{HTML}{C0C0C0}

\newcommand{\hpcasubmissionnumber}{NaN}
\title{Maia 200: A Software Defined Dataflow System for Large-scale AI Acceleration}

\newcommand{\hpcapubid}{0000--0000/00\$00.00}
\newcommand\hpcaauthors{Sherry Xu, Marco Heddes, Jackson Peng, Tom Savell, Monica Tang, Prashant Ranjan,\cr
Jesse Benson, Ofer Dekel, Saurabh Dighe, Anupama Kurpad, Artour Levin, Matthew Mattina,\cr
George Petre, Cheng Tang, Yuan Yu, Li Zhang, Torsten Hoefler}
\newcommand\hpcaaffiliation{Microsoft Corporation}
\newcommand\hpcaemail{}

\author{
  \IEEEauthorblockN{\hpcaauthors{}}
    \IEEEauthorblockA{
      \hpcaaffiliation{} \\
      \hpcaemail{}
    }
}

\fancypagestyle{camerareadyfirstpage}{  \fancyhead{}
  \renewcommand{\headrulewidth}{0pt}
  \fancyhead[C]{
    \ifdefined\aeopen
    \parbox[][12mm][t]{13.5cm}{\hpcayear{} IEEE International Symposium on High-Performance Computer Architecture (HPCA)}    
    \else
      \ifdefined\aereviewed
      \parbox[][12mm][t]{13.5cm}{\hpcayear{} IEEE International Symposium on High-Performance Computer Architecture (HPCA)}
      \else
      \ifdefined\aereproduced
      \parbox[][12mm][t]{13.5cm}{\hpcayear{} IEEE International Symposium on High-Performance Computer Architecture (HPCA)}
      \else
      \parbox[][0mm][t]{13.5cm}{\hpcayear{} IEEE International Symposium on High-Performance Computer Architecture (HPCA)}
    \fi 
    \fi 
    \fi 
    \ifdefined\aeopen 
      \includegraphics[width=12mm,height=12mm]{ae-badges/open-research-objects.pdf}
    \fi 
    \ifdefined\aereviewed
      \includegraphics[width=12mm,height=12mm]{ae-badges/research-objects-reviewed.pdf}
    \fi 
    \ifdefined\aereproduced
      \includegraphics[width=12mm,height=12mm]{ae-badges/results-reproduced.pdf}
    \fi
  }
    \fancyfoot[C]{}
}
\fancyhead{}
\renewcommand{\headrulewidth}{0pt}

\maketitle

\ifdefined\hpcacameraready 
  \thispagestyle{camerareadyfirstpage}
  \pagestyle{empty}
\else
  \thispagestyle{plain}
  \pagestyle{plain}
\fi

\newcommand{\hpcaheight}{0mm}
\ifdefined\eaopen
\renewcommand{\hpcaheight}{12mm}
\fi

\thispagestyle{empty} 
\begin{abstract}
We introduce Maia 200, an advanced AI accelerator delivering high performance—\num{10145} Tflop/s FP4 and \num{5072} Tflop/s FP8 within a 750W TDP and 7 TB/s HBM bandwidth. Maia exemplifies a new class of Software Defined Locally Accessed Dataflow Architectures (SDLA), which explicitly program dataflow engines to orchestrate highly specialized memories and data movement engines. This approach shifts the focus from today's thread-centric to data-movement-centric architecture, improving efficiency and scalability. Our taxonomy of data management, inspired by Flynn’s classification, highlights how SDLA addresses challenges in modern AI computing. Maia 200 achieves significant cost and energy savings while supporting massive parallelism for AI inference workloads, making it a compelling solution for next-generation high-performance computing systems.
\end{abstract}

\section{Introduction}

The advances in modern generative AI techniques have taken the high-performance computing community by storm. AI workloads, especially inference of Large Language Models (LLMs), are now consuming most of the compute cycles world-wide and building efficient high-performance systems to support those workloads is crucial.  We are in the middle of one of the biggest infrastructure buildouts in human history – while training systems are deployed in large datacenter supercomputers with hundreds of thousands of accelerators~\cite{xAI,meta}, inference systems even outnumber their computational capacity. Based on public statements, we conservatively estimate that each day, at least 1.2 trillion tokens  are generated by such systems world-wide. If we assume a relatively small 8 billion parameter (8B) LLM, this would equate to a sustained compute of 6.85 exaflop/s at any given moment; for a larger 400B model, this would be 0.3 zettaflop/s. In addition, the largest supercomputers deployed perform full-scale AI training. Soon, the required total computation will be in the zettaflop/s range and developing TCO- and CO$_2$e-efficient systems is most important. 

Here, we describe Maia 200, a top-of-the-line AI accelerator system, specialized to AI computing workloads delivering best-in-class performance with \num{10145} Tflop/s FP4 / \num{5072} Tflop/s FP8   performance per chip within a 750W TDP (13.3/6.7 Tflop/W) and 7 TiB/s HBM bandwidth. A distributed Maia 200 system integrating \num{6144} chips offers up to 62 exaflop/s FP4 throughput, 43 PiB/s memory, and 8.6 PiB/s Ethernet network bandwidth.  Internal data suggests that Maia 200 saves 30\% cost (TCO) and 15\% energy compared to any other AI accelerator in Microsoft’s fleet due to an aggressive co-design strategy of AI workloads and architecture concepts as well as specific microarchitecture implementation, software stack, and workload implementation without over-specializing to specific problems. 

\textbf{Maia implements a new type of accelerator, which defines a family of emerging architectures that we call Software Defined Locally Accessed Dataflow Architectures (SDLA). Specifically, Software-defined Dataflow uses an abstract machine model that makes data path programming explicit.} It is a similar architectural step to what Single Instruction Multiple Threads (SIMT) was to define the warp-based thread scheduling mechanism establishing classical GPU architectures. It is very different in philosophy as SDLA uses parallel and independent control and data instruction streams to efficiently orchestrate highly specialized memories and data movement engines, \textbf{taking the spotlight away from threads to focus on a data-movement centric view with localized data access.}

We first describe the fundamentals of Software-defined Dataflow and its variant SDLA and then dive into details on the Maia 200 workload requirements, chip and system architectures.

\begin{figure}[t]
	\centering
	\includegraphics[page=1, width=\columnwidth]{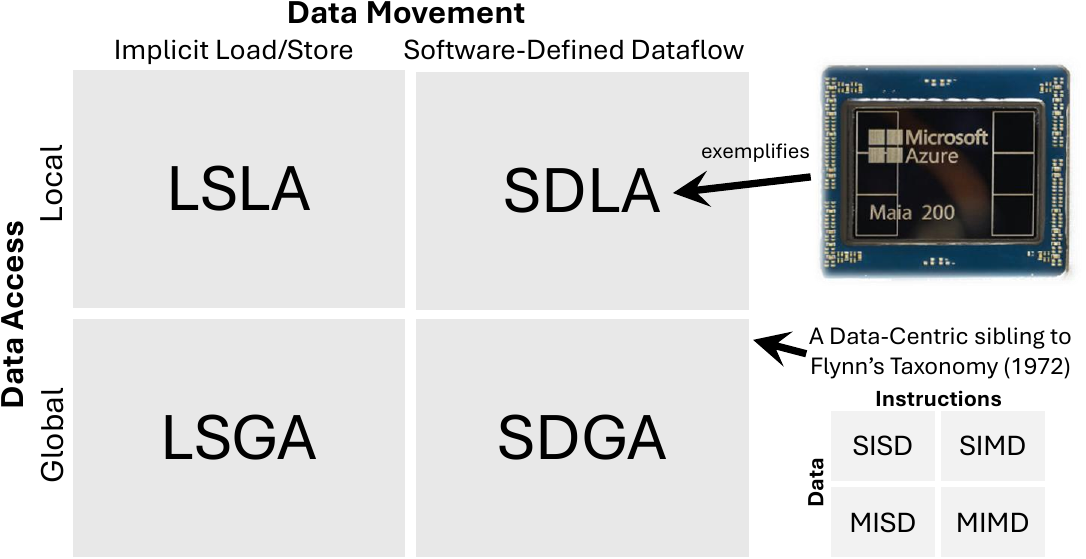}
	\caption{A Modern Data Management Classification of Processors Inspired by Flynn's Original Instruction-focused Taxonomy from 1972.}
	\label{fig:sdda}
\end{figure}

\section{Accelerating Data Management with Software-Defined Dataflow}

We describe a new family of computer architectures that is aimed specifically at accelerating AI workloads at scale. The main optimization goals are to optimize TCO and lower CO$_2$e emissions by improving silicon area and overall system and energy efficiency. We recognize that data management, which comprises storing, moving, and converting data, is the most challenging problem in modern computer architecture~\cite{DMIAYN,memwall1,memwall2}. Furthermore, AI workloads typically process large amounts of data with massive parallelism in relatively regular coarse-grained blocks with predictable control- and dataflow, which we will later define more formally using the concept of data obliviousness. 

To describe the principles in more detail, we define a data-management taxonomy that is similar to Flynn’s taxonomy of instruction parallelism~\cite{flynn1972some}. Our taxonomy is shown in Fig.~\ref{fig:sdda} and distinguishes two dimensions of how compute elements access data and how data movement is defined. 

The vertical data access dimension is similar to Flynn’s data dimension in that it defines whether the compute units can access only local data (logically distributed memories) or all data is accessed globally (logically a single memory). We call those \emph{locally and globally accessed}. Our horizontal dimension, the data movement dimension is similar to Flynn’s instruction dimension in that it defines how data movement is orchestrated – either as part of the computation instruction stream (Load/Store) or in a separate control instruction stream (which we call \emph{Software Defined dataflow}). 

All four quadrants of our taxonomy have example systems today: LSGA is used in standard multicore CPUs forming the most established shared memory abstraction~\cite{pram}. GPUs or Cerebras’ WSEs~\cite{wse} use LSLA with distributed scratch pad memories. Some CPUs extend the standard load/store model to SDGA by adding programmable DMA units such as Intel’s I/O Acceleration Technology~\cite{ioat} or Data Streaming Accelerator~\cite{dsa}, AMD’s DirectPath I/O~\cite{amd2023iommu}, or ARM’s AMBA DMA~\cite{armamba}; all those use global data access in shared memory. 

In this work, we emphasize Software-Defined Locally Addressed (SDLA) Dataflow architectures. Specifically, SDLA separates dataflow execution from control to allow the programmer to orchestrate details of data management in a fine-grained, parallel, and asynchronous manner to enable ideal overlapping of computation and data movement. Furthermore, SDLA’s distributed memories provide significant opportunities for specialization and acceleration such as right-sized memories collocated with execution units and optimized data movement and type conversion acceleration units. Some existing architectures such as AMD’s Xilinx-based Distributed Network Architecture (XDNA)~\cite{xdna} or Samba Nova’s Reconfigurable Dataflow Architecture~\cite{sambanova} would fall roughly into the SDLA category and industry adoption is growing. In this work, we focus on our Maia 200 architecture, the second-generation SDLA from Microsoft. 

SDLA defines a new abstract machine model as a contract between programmers and computer architects. It can be seen as one step beyond Single Instruction Multiple Threads (SIMT) architectures that form the basis of classical GPU acceleration~\cite{simt}. SDLA differs in that it does not focus on sharing instruction streams between threads but on the efficiency of data management. While SDLA enables instruction stream sharing between processing elements, it does not enforce full warp-style synchronization at the program level, enabling independent programming of all units. SDLA fosters efficient programming of spatial architectures where data locality and management are first-class citizens~\cite{spatial1,spatial2,patspatial}.

Fundamentally, SDLA is based on three foundational principles from a programming perspective:
\begin{enumerate}
	\item \textbf{SDLA exposes parallel low-level control programming of accelerated data movement, conversion, and synchronization engines to the programmer, compiler, and software stack.} This enables highest performance and highest power efficiency through explicit control of data management.
   \item \textbf{SDLA allows programmers to manage an efficient system of distributed and specialized (on-chip and in-pod) memories.} This enables architects to optimize memories and accelerate data movement with special engines and programmers to minimize overheads.
   \item \textbf{SDLA enables architects to build systems offering mostly deterministic performance to enable optimized scheduling.} This is powerful for implementing kernels such as transformer attention, where much effort is spent to adopt code to idiosyncrasies of current accelerators~\cite{fa3}.
\end{enumerate}

\textbf{These principles enable compiler engineers to build emulation layers and scheduling libraries} that behave similarly to current firmware in existing accelerators and provide a simplified programming interface. Yet it empowers expert programmers to either use high-level programming for non-critical pieces or dive into every detail of memory management and scheduling for performance-critical code pieces. Current classical accelerators like GPUs only offer one level of programming such as CUDA while scheduling (e.g., warps) and fine-grained DMA control are implemented in hardware or firmware and thus out of the reach of typical programmers.

\textbf{These principles also enable computer architects to co-design execution engines for their tasks} such as data movement, synchronization, or computation and thus optimize silicon design. Current expert GPU programmers use “warp specialization” as a technique to pipeline memory accesses with computations. Yet, this software technique uses standard warp threads for both purposes. SDLA is one step ahead in that execution engines are specialized to their purpose and thus more efficient when co-designed in hardware. 

In summary, \textbf{SDLA redefines the contract between architects and programmers to provide lowest level control but at the same time enable an advanced software stack and compiler to hide the complexity.} Similar to how classical GPUs hide details of the instruction sets (ISAs) through LLVM-like PTX abstractions~\cite{ptx}, SDLA enables two layers of programming where the programmer can choose to write assembly code (cf. Streaming Assembly, SASS~\cite{sass}) or higher-level CUDA-like code. The \textbf{control path} (processors) is programmed in a standard low-level systems language such as C or C++ while the \textbf{programmable data path uses a dataflow instruction set architecture} to execute complex dataflow dependencies. The software stack can then build higher-level functionality such as Python/PyTorch or Triton~\cite{triton} interfaces on top of the control and programmable data paths.

At a more abstract level, SDLA extends the classical GPU’s SIMD to SIMT transition with a focus on data management. It provides access to the programmable data path architecture details for explicit accelerated data movement between specialized memories on-chip and across chips. Software-defined means that the application logic is supported by \textbf{control programs that orchestrate the data path}. This makes SDLA  a statement about control (synchronization and data movement) and not about application logic and it thus supports both SPMD (e.g., OpenMP~\cite{openmp}, MPI~\cite{mpi}) and MPMD (e.g., separating different coupled components such as prefill and generation phases in LLM inference) programming styles.

\begin{figure}[t]
	\centering
	\includegraphics[page=2, width=\columnwidth]{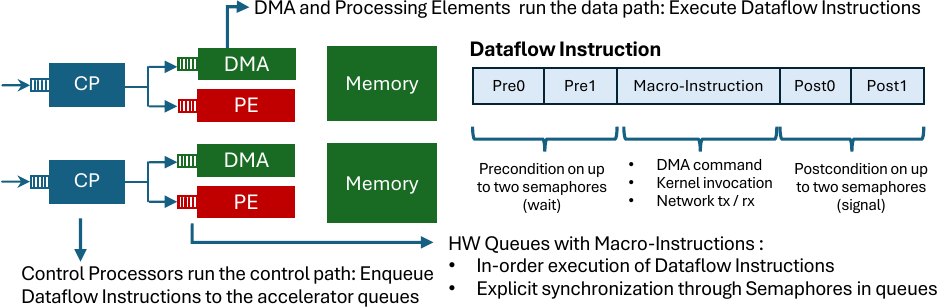}
	\caption{Abstract Machine Model for our Dataflow ISA.}
	\label{fig:amm}
\end{figure}

We show an example of our generic SDLA dataflow architecture in Fig.~\ref{fig:amm}. It uses control programs written in a programmer-accessible high-level systems language such as C or C++. Those programs run in the control path and orchestrate the dataflow computation in the data path through queued instructions. The data path itself comprises different units for data management and processing. Those units support a dataflow instruction set architecture (Dataflow ISA) that defines synchronization pre-conditions, a macro-instruction, and synchronization post-conditions. The management of pre- and post-conditions can be fully implemented in hardware and macro-instructions can either be complex hardware-instructions such as a tensor-core matrix multiplication or software functions invoked on programmable processors such as vector units. 

\section{The Maia 200 System}

Microsoft’s AI Architecture (Maia) 200 is Microsoft’s second- generation of AI accelerators implementing an SDLA dataflow architecture. It is highly optimized for Microsoft’s inference workloads to enable extreme-scale transformer-based LLMs at competitive cost and energy consumption throughout the worldwide fleet. It’s designed to run a very defined workload through extreme co-design of hardware, software, deployment, and operational aspects of the system. To achieve this, it embraces a new programming philosophy that delivers lower TCO and energy consumption. Yet it is not overspecialized and supports future workload shifts.  

Maia 200’s architecture takes advantage of the SDLA programming view to improve silicon and energy efficiency and thus cost. Specifically:
\begin{itemize}
	\item Maia 200’s architecture uses specialized small memories attached to specific functional units to improve silicon and energy efficiency.
    \item Maia 200 organizes those memories and compute units hierarchically to take advantage of locality in workloads and programming.
    \item Maia 200 extends seamlessly into the network. The network interfaces are driven as part of the programmable data path, e.g., data moves seamlessly from local SRAM or HBM into remote SRAM or HBM.
\end{itemize}

Maia 200 is fabricated in TSMC’s 3nm process with more than 140 billion transistors on a near reticle-sized monolithic die (26x33mm). It uses CoWoS-S packaging technology to co-locate HBM on a silicon interposer in a 75mm x 75mm package with a total of 750 W SoC TDP distributed via 19 metal layers. It is a full system including tray, rack, and network architecture scalable to thousands of accelerators in a single cluster and it is in production in the fleet today. It is mainly optimized for delivering highest efficiency for massive inference workloads using trillion-parameter frontier models as this is the main demand today.

\subsection{Inference workload challenges at scale}

Maia 200 is specifically designed for inference workloads of large LLMs including Mixture of Expert layers~\cite{moe}. Much of the workload has soft real-time requirements in that a user is waiting for the output relying on relatively strict SLAs for example \num{300} – \num{4000} ms to first token and 20-30 ms per token. Inference generally can be split in two components: prefill and decode. Prefill computes the first output token from the concatenated system and user prompts, which can be tens of thousands of tokens while decode generates token-by-token based on the contextual history in the KV cache~\cite{kvcache}. Thus, prefill usually uses the system architecture in a balanced manner while decode is often memory-bandwidth limited due to the KV cache accesses. 

Batching can change the balance by re-using weights across elements of a batch but decode still requires loading a separate KV cache for each batch entry. Maximal batch sizes are limited by the SLA as well as the accelerator’s memory size and parallelization~\cite{ddl}. Inference systems can utilize lower-precision datatypes such as FP4 or FP8 if used with fine-grained scaling factors~\cite{mxfp,nvfp}. Prefill and decode phases can be disaggregated to bandwidth-optimized and compute-optimized accelerator configurations. Typically, batching results in a relatively large number of compute and communication operations, often involving several kilobytes or megabytes of data.

In general, inference of large LLMs requires a cluster of petaflop/s accelerators to perform the task for multiple requests simultaneously~\cite{vllm,sglang}. The Maia 200 system architecture designs such a supercomputer cluster system with 62 exaflop/s FP4 performance that enables specialization to the different phases and 8.6 PiB/s network connectivity to enable efficient inference.

Alongside those common challenges in large-scale LLM inference, we explicitly model new trends that will shape future workloads:
\begin{enumerate}
    \item Massively growing context windows are required by reasoning models~\cite{rlm} that implement chain, tree, or graph of thoughts, multimodal models including images, voice, and even videos, and retrieval augmented generation (RAG). Those require hundreds of thousands to millions of tokens and extreme bandwidths in wide data paths (Section~\ref{sec:bandwidth}).
    \item Growing models require more compute power and memory bandwidth per token; diffusion models need even more compute power for many iterations (Section~\ref{sec:tvp}).
	\item Coarse-grained sparsity and dynamic computations generate more data movement that can only be scheduled at runtime. The prime examples are Mixture of Experts (MoE)~\cite{moe} systems that often require all-to-all communication at extremely high bandwidths. Thus, the control path must be fast to implement data-dependent control (Section~\ref{sec:swdataflow}). 
    \item Agentic systems and tool use agents cause more complex workflows~\cite{agentic}, potentially involving multiple accelerators with different models that need to be tightly coupled (Section~\ref{sec:anc}).
\end{enumerate}
In general, the extremely fast-growing demand makes reducing energy consumption and cost even more important in the near future.

\subsection{The Maia 200 System on Chip}

While Maia’s foundational innovation lies in adopting the SDLA principles outlined above, the Maia 200 SoC requires many staple features for modern AI accelerators, such as: 

\textit{Specialized (scaled block) narrow datatypes to enable hardware-efficient inference and training without significant accuracy loss:} Maia 200 emphasizes block-scaled FP4 and FP8 tensor compute.

\textit{A capable Network on Chip (NoC) to connect the reticle-sized massive monolithic chip:} Maia 200’s NoC is physically split into a logical high-bandwidth data NoC and a specialized control NoC to keep data- and control traffic separate. The data NoC also supports QoS to prioritize different data traffic (e.g., coming from NIC or HBM).

\textit{High HBM bandwidth, connected to the NoC to deliver highest bandwidth to all units:} Maia 200 uses six HBM3e stacks.

\textit{Tensor cores with complex operations:} Maia 200’s flexible Tile Tensor Units support matrix multiplications as well as convolutions natively, a feature only announced for future GPUs.

\textit{Advanced power management features (DVFS):} Maia 200 offers power steering through separate clock domains for compute, the NoC, and the HBM subsystem. This allows us to specialize chips at deployment in different configurations, e.g., optimized for prefill or generation.

In addition to those staple features that are a must-have for any modern AI accelerator, Maia 200 innovates on multiple fronts using SDLA principles:
\begin{itemize}
	\item Control and data management are implemented separately, allowing us to optimize highly-specialized units such as DMA, synchronization (hardware semaphore), and processing elements.
	\item The NoC offers multicast capability to efficiently distribute data to all clusters at full bandwidth.
	\item PEs are equipped with highly specialized memories, for example, the Tile Tensor Unit uses highly specialized memories to access inputs and outputs.
	\item Additionally, an on-chip hierarchy of specialized memories supports efficient near- (within an on-chip cluster) and far (remote on-chip cluster) communications.
	\item The system offers several specialized programmable data-movement and data management accelerators supporting: Strided DMA operations for 1D, 2D, 3D, 4D tensors including scatter/gather support and line-rate datatype casting and sparsity support in the Tile DMA engines.  
	\item Extreme network bandwidth while keeping the network below 20\% of the system cost, key are the 28 integrated co-designed NICs with a simplified transport protocol and an innovative network topology using mostly fixed links saving switches and cables. 
\end{itemize}

\begin{figure}[t]
	\centering
	\includegraphics[page=3, width=\columnwidth]{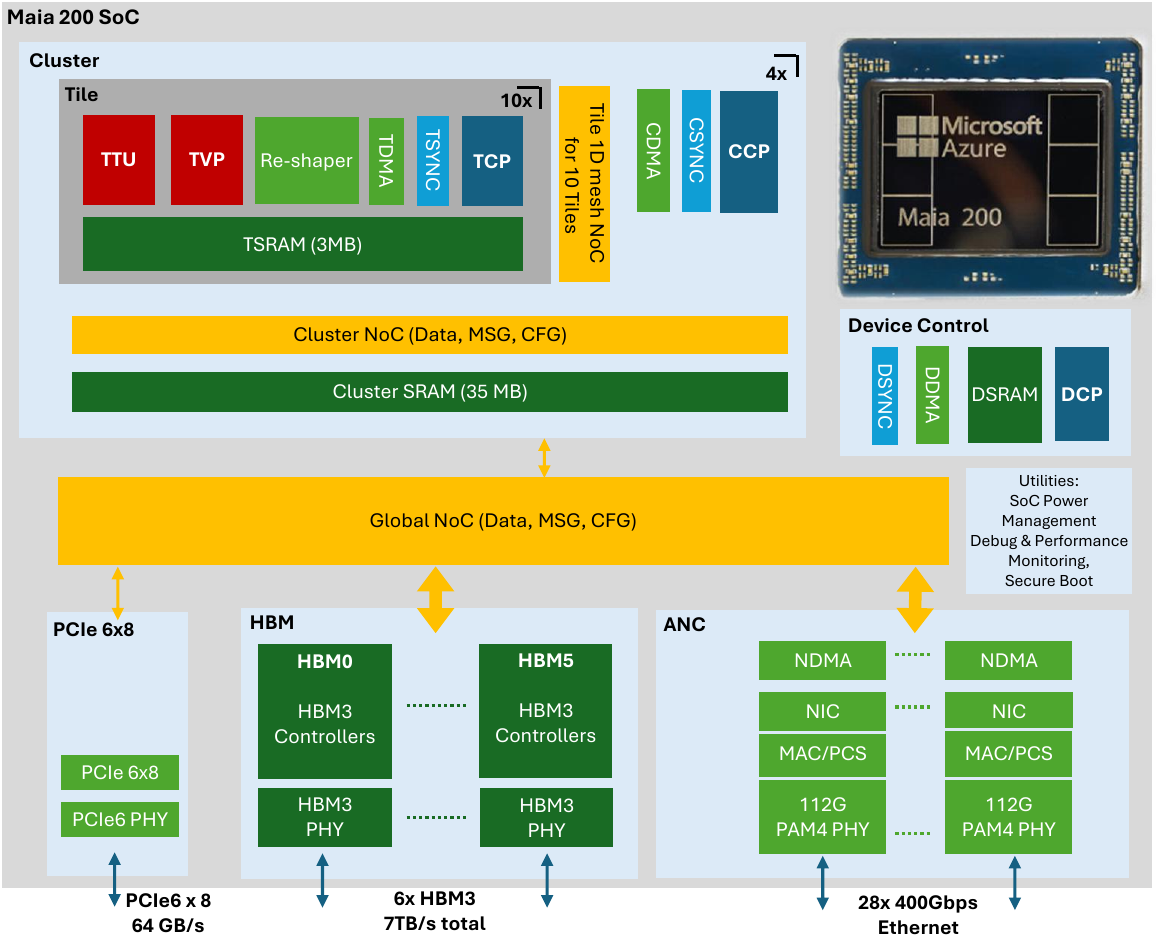}
	\caption{Maia 200 Abstract System on Chip Architecture. Processing blocks are shown in red, the data path is green with data movement accelerators in light green and memories in darker green, the NoC is in yellow, and the control path is in blue with synchronization engines in light blue and control processors in darker blue.}
	\label{fig:soc}
\end{figure}

Fig.~\ref{fig:soc} shows the overall logical architecture of a Maia 200 SoC. The central compute complex is formed by four Clusters on the chip die. Each of the four clusters contains nine or ten Tiles. Each Tile contains one Tile Tensor Unit (TTU) and one Tile Vector Processor (TVP), a specialized tile memory, DMA engines, Synchronization (Sync) engines, and a Tile Control Processor (TCP). The tiles are connected by a special Tile 1D mesh NoC to enable faster communication among each other. Each of the four Clusters has a separate Cluster SRAM (CSRAM), a Cluster Control Processor (CCP), Synchronization blocks, and DMA engines. The global NoC (GNoC) connects all clusters, the HBM, the host system via PCIe, and the Integrated NIC block (AI Network Controller, ANC). The chip is coordinated by a Device Control block including a Device Control Processor (DCP) as well as SRAM and specialized DMA and Sync engines. The SoC also contains utility functions for power management, performance monitoring, debugging, and secure boot. All-in-all, Maia 200 is a capable microarchitecture implementing the SDLA ideas. We will elaborate on some of the design decisions below.

\subsubsection{Explicit Scratchpad versus Caches for SoC SRAM}

There is a subtle tradeoff between caches and explicitly programmed scratchpad memories. Hiding fast cache memories between the main memory and registers is simplest for entry-level programmers. However, for extreme performance-conscious ninja programming, the discussion becomes more complex. Ninja programmers usually consider the details of the cache to optimize their data access patterns. This can be quite complex and non-portable depending on the complexity and documentation of the cache itself. For example, cache replacement strategies, associativity, and cache-line sizes must be considered when writing the highest-performing code~\cite{cache1,cache2,cache3,cache4,cache5,cache6}. Yet not all code pieces are performance critical and getting reasonable performance at lowest initial investment and simpler programming remains an important goal.

First, we look at AI workloads: An important property of algorithms is whether they are data oblivious. An algorithm is data oblivious (or oblivious for short) if all memory accesses and dataflow can be computed based on a small number of parameters that are available at compilation time. Most AI workloads are a static composition of fixed kernels (MMM, SoftMax, non-linearities, etc.) that are all oblivious, making the overall dataflow oblivious and thus plannable~\cite{oblivious}. There are some small exceptions such as continual/dynamic batching, Mixture of Experts,  other forms of sparsity, and early termination~\cite{sparsity}. All those can either be handled in a parametric way that takes advantage of their near obliviousness, or they are at the very coarse control level and thus not critical for performance. In summary, the \textbf{mostly oblivious and thus plannable AI workloads enable ninja programmers and compilers to statically plan all data movement and placement in an overall optimized (if not optimal) schedule using distributed memories and accelerated data paths efficiently}~\cite{map1,map2,map3,patspatial}.

Caches are considered vital for many applications and enable extremely quick time to first implementation. In fact, one can show that, for an LRU replacement cache with unit-sized cache-lines that is twice as large as a scratchpad memory using the optimal (offline) memory management scheme, the cache will at most have twice as many misses~\cite{scache,snir}. Thus, LRU caches are only a small constant factor of about four away from optimal. This minor overhead is often tolerable, so CPUs and most general-purpose GPUs use caches to make programming easier. Thus, we carefully study the benefits programmers as well as computer architects could gain from replacing caches. 

Alongside the previously mentioned energy and cost savings—up to four times—ninja programmers are also able to orchestrate highly efficient data transfers for AI tasks. Similarly, for a set of HPC workloads, including many non-oblivious codes, Marinelli et al.~\cite{compad} showed a mean geometric speedup of 13\% when manually porting those from caches to scratchpad memories. Specifically, ninja programmers benefit from scratchpad memories in that 
(1) Data can be placed and packed at word-granularity in all memories. Cache line size and cache line sharing play no role, which simplifies the handling of non-contiguous fine-grained data significantly. 
(2) Programmers do not need to worry about replacement schemes and keeping the right data in cache or issuing special uncached load/store instructions for streaming accesses (such as weights) that may block the pipeline. They simply keep the data that they want in scratchpad and overwrite it explicitly.
(3) Programmers do not need to account for hidden prefetch streams. In SDLA, programmers orchestrate all details of asynchronous pipelining through accelerated data movement engines.
And finally, (4) specific extensions such as 1D/2D/3D tensor loads and/or datatype conversions can save a significant number of CPU instructions~\cite{ssr,stash}.

Hardware implementations of caches are also more expensive than software: in addition to the SRAM cells, they require tag memory arrays and address remapping logic. The area and energy overhead for this logic is around 30-35\%~\cite{compad,banakar}. Furthermore, the access latency is increased by 10-15\% due to address decoding and mapping~\cite{compad,banakar}. Overall, this causes an area-time and dynamic energy overhead of up to 43\% in practice~\cite{compad}.

Our explicitly programmed SDLA memory architecture enables Maia 200 to spend less than 20\% of its chip space on memory, which is far less than many CPUs with smaller cache sizes. Furthermore, all SRAMs and the HBM are ECC protected and can be scrubbed on demand to protect the large die from bit flip corruption. The specialized DMA engines to orchestrate the data movement occupy only negligible die area.

To unify the best of hardware design and plannable software, Maia 200 aims to use compilers to utilize software-managed scratchpad memories at low programmer complexity and reasonable performance~\cite{compiler1,compiler2}, while ninja programmers will have the choice to orchestrate the data movement at the lowest levels. Compilers offer basic support for Python/PyTorch and Triton programming at the top level while ninja programmers can dive into writing C/C++ control programs for some kernels to use the architecture at highest efficiency. In this work, we focus on both the hardware and system architecture and thus will not describe the software stack.

\subsubsection{Dataflow Programming and Instructions}

Maia 200 combines several types of computational and data management (memory and data movement) units to implement an SDLA microarchitecture. It combines two main types of compute units paired in each Tile: A Tile Tensor Unit (TTU) and a Tile Vector Processor (TVP). TVPs can execute complex programs written in C/C++ while the TTU supports a large set of fixed dataflow instructions (matrix multiply and convolutions). DMA engines support a fixed instruction set for moving data between memories. Each Tile DMA engine includes a Reshaper engine that can perform data conversions between different formats and data layout changes such as transposition at line-rate.
 
Each of those engines supports a specific set of powerful \textbf{macro-instructions}, which can be full programs (e.g., in TVPs) or fixed functions (e.g., in TTUs or DMA engines). Those macro-instructions are part of a \textbf{dataflow instruction set architecture (DISA), which defines a set of dataflow preconditions, a macro-instruction to invoke, and a set of dataflow postconditions} as shown in Fig.~\ref{fig:amm}. Each basic (dataflow and compute) execution engine supports DISA instructions that are used to orchestrate the overall dataflow program. Dataflow conditions are managed using sync engines that offer logical semaphore objects that are assigned by the control program. Each semaphore can be configured to fire at a specific threshold. Each dataflow instruction can be configured to wait for up to two semaphores before executing (precondition). After the execution finishes, it can signal up to two semaphores (postconditions), which in turn could trigger another dataflow macro-instruction. All wait and signal conditions have a programmable decrement or increment value and commands can be chained to enable more complex synchronization conditions.

Maia 200 includes a hierarchy of different types of control processors that orchestrate the dataflow programs using the execution units and sync engines to manage the overall data flow. The control program orchestrates the setup of the dataflow instructions and semaphores ahead of time and the overall dataflow program executes asynchronously with respect to the control program (which usually runs ahead). This scheme can implement optimal dataflow for oblivious programs without any delays due to the parallelism of control and dataflow in hardware. We offer a simple C++-based low level programming language called NEsted PArallel Language (NEPAL) to program the control path. 

\subsubsection{Connecting Memories and Compute Engines}\label{sec:bandwidth}

Maia 200 uses memories that are directly attached to compute engines for highest efficiency. They are sized according to Little’s Law~\cite{little} to satisfy the dataflow throughput of the engines for pipelined and overlapped loading and computing as well as to maximize reuse of data and minimize data movement and addressing overheads.

Each tile has 3 MiB of specialized SRAM to hold the operands for the TTU. Each TTU can perform \num{65536} Multiply Accumulate (MAC) FP4 operations per cycle. It reads two input matrices of size 32x64 and outputs, or accumulates into a 32x32 FP32 or BF16/FP16 matrix. For FP8 and BF16, it performs half or 1/8th of the MAC operations and supports 32x32 or 32x8 input matrices, respectively. This means that the input dataflow bandwidth at 2 GHz is \num{2024} GiB/s per operand and the output bandwidth is up to \num{4096} GiB/s (with an additional \num{4096} GiB/s read bandwidth for accumulation). Thus, the memories can hold work for the TTU for more than 700 cycles, which enables full-utilization pipelining. Each TVP works on TSMEM such that it can perform post-processing of the matrix outputs and TTUs can also read one input operand from TSMEM to enable efficient chains of operations. The memories are orchestrated through Tile DMA engine dataflow commands, which support optional reshaping. The Tile DMA can use the Tile NoC to communicate with neighboring tiles or the Cluster NoC to access the Cluster SRAM.

The overall system comprises many dozens of different DMA engines, some move data between on-chip SRAMs and/or HBM, some move data into the network to remote SRAMs or HBM. NoC QoS priorities can be assigned on a per-command basis. The utility fabric fully connects all Sync blocks, which can each notify any other Sync block of the SoC at lowest latency.

\subsubsection{Datatypes and Conversion Acceleration}\label{sec:tvp}

Maia 200 supports a flurry of different datatypes and formats. In line with its dataflow principles, it separates between \emph{storage data types and compute data types}. Storage data types are used in the data management path to reduce the number of bits transported and stored while maintaining better accuracy than lower bit formats. Sparse tensor storage is the main storage datatype. 
The Reshaper unit can convert between wider and narrower data types using stochastic rounding or round to nearest. It can also load from 8:16 or 4:16 sparse tensors to/from dense tensors at full bandwidth.  

Each TTU supports BF16, FP8, FP6, and FP4, with \num{8192}, \num{32768}, \num{32768}, and \num{65536} MACs, respectively. FP6 achieves the same performance as FP8 but at significant energy savings through gating. TTUs read matrices of size  32xKx32
\begin{wrapfigure}{r}{0.24\textwidth}
	\centering\small
	\vspace{-6pt}
	\begin{tabular}{c | c}
						
						\rowcolor{tablegray}
				\textbf{Datatype} & \textbf{Lanes} \\
		
				\hline

		INT8/UINT8 & 512 \\
		INT16/UINT16 & 256 \\
		INT32/UINT32 & 128 \\
		BFP16 & 256 \\
		FP16 & 256 \\
		FP4 (E2M1, E1M2) & 256 \\
		FP6 (E3M2, E2M3) & 256 \\
		FP8 (E4M3, E5M2) & 256 \\
		FP32 & 128\\

	\end{tabular}
	\vspace{-7pt}
		\end{wrapfigure}
in blocked FP4 and FP8 formats (e.g., K=32 for FP8 and K=64 for FP4) and output FP32 or BF16 at 262.14 Tflop/s for FP4 multiplications at 2 GHz. Each TVP supports a richer set of data formats for vector operations, which are shown in the adjacent table. TVPs implement 256 lanes for types up to 16-bit width at 3.07 Tflop/s and 128 lanes for FP32 at 1.54 Tflop/s at 2 GHz.

The FP8 format supports E4M3 and E5M2 layouts, the FP6 format supports E2M3 and E3M2 layouts and the FP4 format supports the E1M2, E2M1 layouts. Maia 200 allows programmers to compose those formats into OCP compliant MXFP data types with E8M0 scaling factors and group sizes of 32~\cite{mxfp}.

\subsubsection{Software-Defined Dataflow Control}\label{sec:swdataflow}

The control path is the last missing piece to a full SDLA microarchitecture. Maia 200 combines three hierarchy levels of control: the SoC, Cluster, and Tile levels. Each of those levels has several control processors. Those processors are running C/C++ programs of the user to orchestrate the data path using the Dataflow ISA defined above. Those processors usually run slightly ahead of the dataflow to configure the engines before they are used. Slightly larger staging memories (CSRAM, TSRAM) support this asynchronicity and allow the control processors to stash enough work such that units are never waiting for work. Data-dependent data-to-control dependencies require a quick connection from the data path to the control path.  

Maia’s hierarchical architecture benefits such data-dependent control in that they can happen in the same tile and the data-to-control remains between adjacent units. For example, for a mixture of experts operation in which a routing function decides where to send activations to, the TVP evaluates the output of the TTU routing matrix multiplication in TSRAM and hands off the result to the TCP in a small number of cycles to kick-off preconfigured data movement operations to each destination. The control for this data-dependent flow (from TTU result to the TVP to the TCP) can be set up in advance through semaphores and DMA operations that get their scatter-gather operations from the TVP. This enables fastest local control for quickly changing data-dependencies. Similar methods can be used to support local implementations for sparse processing~\cite{sparsity}.

\begin{figure}[t]
	\centering
	\includegraphics[page=4, width=\columnwidth]{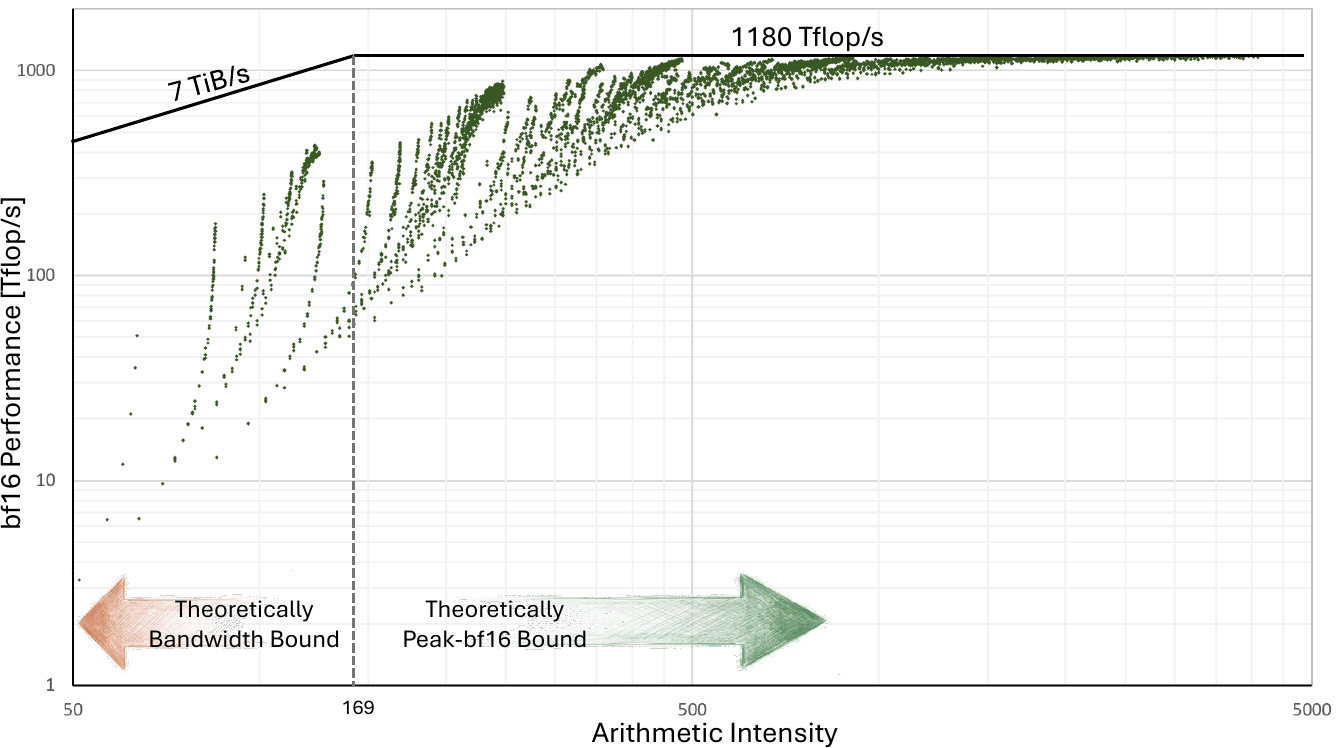}
	\caption{BF16 Matrix Multiply Performance for 6143 Relevant Matrix Sizes.}
	\label{fig:bf16}
\end{figure}

\subsubsection{Power, Energy, and Management}

Maia 200 is a complex system, and we skipped over most of the management functionality. For example, the details of how each control processor is configured and how kernels are launched are complex but less scientifically interesting. We now highlight some of those aspects that are most interesting to performance-conscious users and engineers.

Maia 200 offers a flexible Dynamic Voltage and Frequency Scaling (DVFS) system where each of the four clusters as well as the Global NoC are in different frequency domains. This allows us to deploy configurations optimized for prefill (high compute and high memory clocks) or token generation (lower compute and high memory clocks) to optimize for specific use-cases.
 
Maia 200 also offers a fully secure boot system and it is seamlessly integrated into Azure’s compute philosophy and management. 

\subsubsection{Roofline Matrix Multiply Performance}

We continue by investigating the most important workload in deep learning inference: low precision matrix multiplication. Here, we focus on the most relevant formats today: multiplying two bf16 matrices and multiplying two fp8 matrices. Both of those operations output bf16 matrices. We benchmark a Maia 200 system with 9 Tiles per cluster enabled running at 2 GHz (unthrottled). Thus, the peak bf16 performance is 262.14 $\cdot$ 4 $\cdot$ 9 = 1180 Tflop/s and the peak fp8 performance is 4785 Tflop/s. The roofline’s ridge point~\cite{roofline} lies at an arithmetic intensity of 674 for fp8 and 169 for bf16, respectively.

\begin{figure}[t]
	\centering
	\includegraphics[page=5, width=\columnwidth]{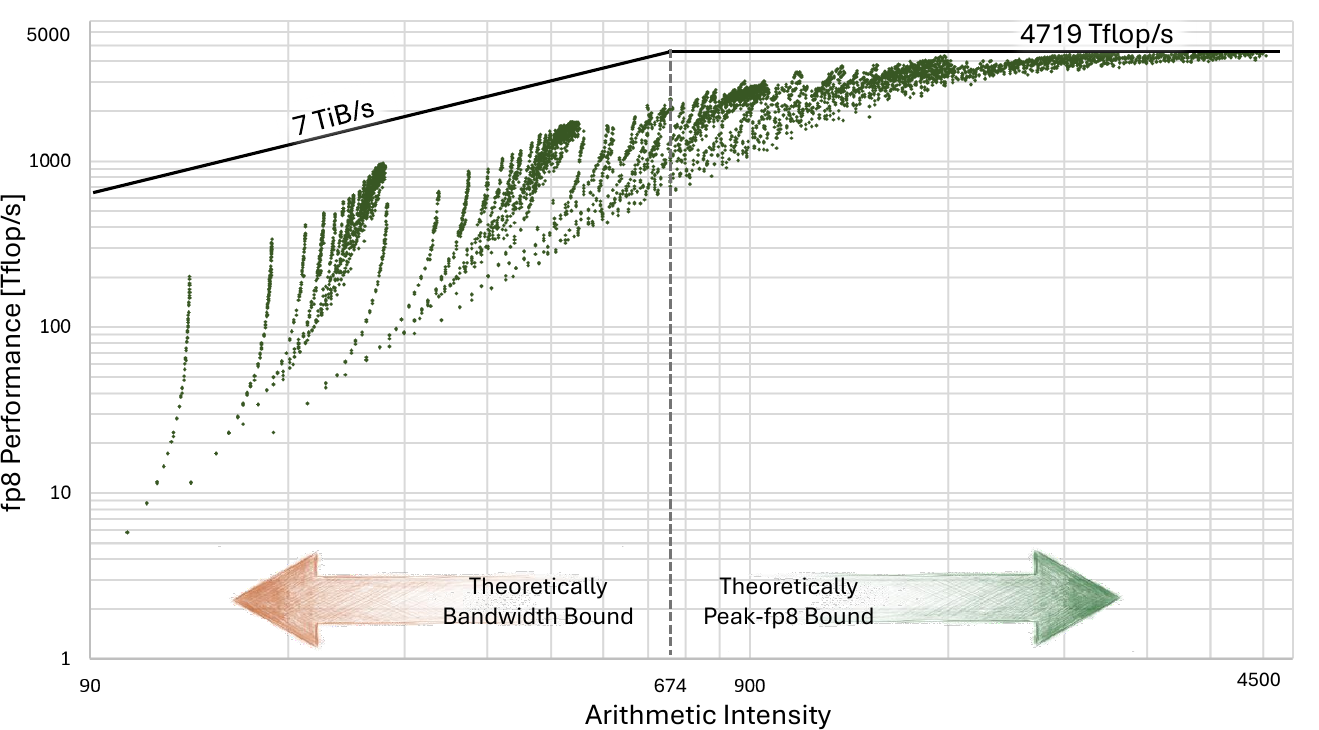}
	\caption{FP8 Matrix Multiply Performance for 6143 Relevant Matrix Sizes.}
	\label{fig:fp8}
\end{figure}

Fig.~\ref{fig:bf16} shows a scatterplot of 6143 relevant matrix multiplications of different sizes that are relevant for AI inference workloads, spanning arithmetic intensities from 85 to 5000. We measured each matrix size performance by repeatedly loading the input matrices from HBM and writing back into HBM. The measurement was repeated until the time exceeded 1ms to amortize the timer accuracy. The variation across measurements is minimal due to the explicit SDLA programming and the absence of caches. We achieve remarkably high \textbf{bf16 roofline efficiency up to 99.69\% of peak} in the compute bounded regime. The near-100\% utilization is due to the asynchronous nature of data movement and SDLA programming. All instructions and data loading can overlap and all units are busy, synchronized by semaphores. We exceed 90\% of the peak flop performance for  multiplications using more than 58 Tflop of compute. In the memory bound regime, we achieve up to 51.4\% of the peak bandwidth due to the smaller size of the matrices (larger ones quickly become compute bound). We achieved higher than 50\% bandwidth for combined input and output operands bigger than 113.5 MiB. In both regimes, we outperform comparable GPUs and CPUs but consciously refrain from a direct comparison as this study aims to show the architectural efficiency of SDLA and Maia 200 using real benchmarks.

Fig.~\ref{fig:fp8} shows similar data and performance for fp8 matrix multiplications in a very similar setup with the same matrix sizes and experimental methodology. Here, we span arithmetic intensities from 95 to 4600 and achieve up to 96\% of the peak performance in the compute-bound regime and up to 56\% in the memory-bound regime.

\subsection{Connecting Many SoCs into a Distributed Compute Cluster}\label{sec:anc}

Maia 200 uses 28 integrated 400 Gbps Ethernet-based AI Network Controllers (ANC) for a total bandwidth of 1.4 TB/s full duplex. The ANCs are heavily optimized for minimal chip space and energy consumption. They are fully integrated into the SDLA microarchitecture to which they offer programmable DMA engines, synchronization, and seamless data movement to remote SRAM and HBM memories. Over the lossless (PFC) Ethernet L2 network they run Microsoft’s in-house AI Transport Layer version 2 (ATLv2) protocol~\cite{patanc} to implement network transport that later influenced the standardization of Ultra Ethernet~\cite{uespec,ue}.

For packet transmission, the ATLv2 protocol uses standard L2 Ethernet frames with L3 IP routing headers and full end-to-end AES-GCM-256 encryption and a simple window-based congestion control scheme. It supports ECMP and entropy vectors encoded in the position of the UDP source port for per-packet load balancing and it uses a protocol similar to REPS~\cite{reps,patreps} for recycling entropies of known-to-be-good paths. It supports selective retransmit to overcome well-known limits of Go-Back-N retransmission~\cite{dc-limits}. Furthermore, it supports QoS not only in the network but also in the GNoC towards HBM to make sure messages are delivered without causing network back-up.

\begin{figure}[t]
	\centering
	\includegraphics[page=6, width=.7\columnwidth]{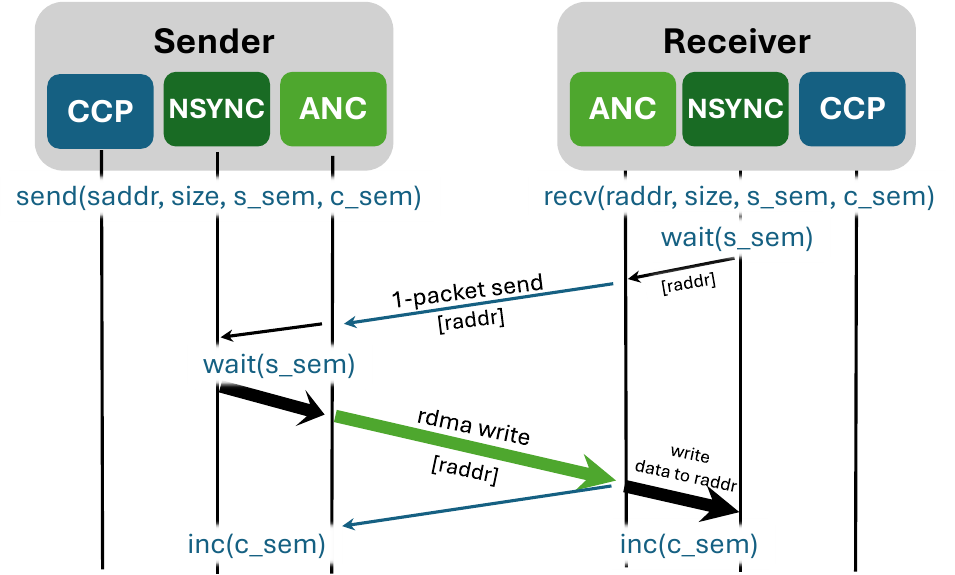}
	\caption{ATL’s Receiver-driven Messaging Scheme.}
	\label{fig:atl}
\end{figure}

The SDLA functionality to access remote memories are exposed through send/receive where the sender specifies a send buffer and the receiver specifies a receive buffer (or multiple in the case of broadcasting) for the message to be copied into. We implement those semantics using Remote Direct Memory Access (RDMA) in the ANC. We use a receiver-driven messaging scheme that is optimized for AI communication in Collective Communication Libraries (*CCLs) to minimize message matching state. This scheme inspired the AI Base profile that was later standardized in Ultra Ethernet~\cite{ue,uespec} and is shown in Fig.~\ref{fig:atl}.

In Fig.~\ref{fig:atl}, the Sender CCP posts a send operation that will start after the start semaphore (s\_sem) fires and it will increment the completion semaphore (c\_sem). At the receiver, the CCP posts a receive with similar start and completion semaphores. The receiver waits for its s\_sem to fire and then sends a control message containing the receive address to the sender. The sender’s ANC will await a message from the receiver and once it receives it wait for the s\_sem to fire. If both conditions are true, it will issue an RDMA write message to the receiver’s address. Upon reception of the write command, the receiver increments its c\_sem and once the sender receives the acknowledgement of the successful write operation, it in turn triggers its local c\_sem. Note that this protocol does not require any buffering in the ANC as it can always retransmit the data via DMA from the source memories. The ANC itself splits each message into multiple packets that are sent in any order and both sender and receiver use bitmaps to track message completion. The system is optimized for messages larger than 4 kiB, which is common in AI workloads~\cite{patanc}.

The system supports packet spraying across multiple ANCs along different network paths. ANCs also support remote semaphore increment together with data transmission as well as remote data broadcast to clusters on the GNoC. Messages can be written into remote CSRAM or HMB. This makes remote nodes seamless peers in the overall SDLA programming model. 

\begin{figure}[t]
	\centering
	\includegraphics[page=7, width=\columnwidth]{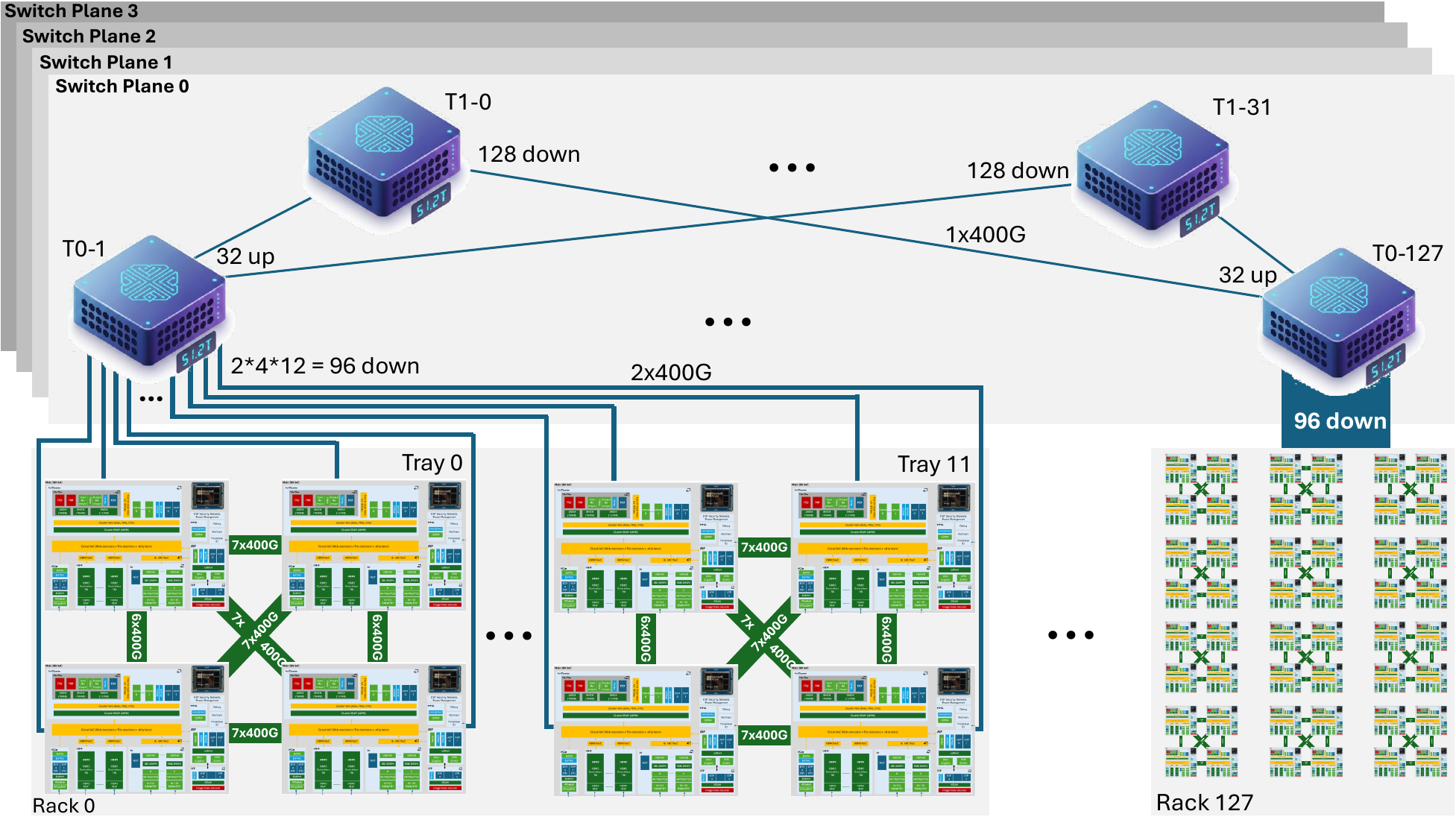}
	\caption{Maia 200's Two-Tier Network Topology.}
	\label{fig:topo}
	\vspace{-10pt}
\end{figure}

The system design combines fixed connections on the same blade with switched connections. The topology is a special case of a 2x2 1D Hamming Mesh~\cite{hxmesh} where cross links are added on each board establishing full connectivity in the group collocated on a physical tray. Of the 28 ANCs per SoC, 20 are connected with fixed links as shown in Fig.~\ref{fig:topo}, and eight are connected to a switched network. The switched network consists of four identical planes and each SoC connects with two 400G ANCs to each plane. While the switched ANC links can be connected into any topology with hundreds of thousands of endpoints, the current design point is aimed at a two-tier network with a maximum of \num{6144} SoCs for inference deployments. In this configuration, each of the 51.2T (128 400G ports) Tier-0 (T0) switches is connected to 48 Maia 200 SoCs in 12 trays with two links each. The remaining 32 ports are connected to up to 32 Tier-1 (T1) switches with a 1:3 oversubscription ratio. Thus, the overall number of SoCs is 48$ \cdot $128 = \num{6144}. Smaller subset configurations are possible and deployed in the field. The chips and trays themselves can also be used to build larger-scale systems by adding more tiers of switches or changing the topology.

\subsubsection{Collective Communication Performance}

Collective communications are critical for deep learning performance. We now discuss how basic algorithms are implemented in Maia 200.  We focus on initial implementations of two algorithms, direct-connect and ring~\cite{colls}, while we fully acknowledge that the performance can be improved with more elaborate algorithms (e.g., Bine Trees~\cite{bine}). Yet, our purpose here is to demonstrate the efficacy and performance of the SLDA architecture and specifically Maia 200’s system design running production workloads. 

The direct connect algorithm simply sends the right pieces of the data directly to the target endpoints. The ring algorithm established a logical one-dimensional ring connecting all chips and sends chunks of data in a pipelined manner along the ring such that all accelerators send and receive at all times. Details on those standard algorithms can be found in Thakur et al.~\cite{colls}. The key difference between them is that the direct connect algorithm has a message depth (longest path of depending message synchronizations) of one: no message depends on another message to be received. Thus, direct connect has the lowest latency. Yet, it may not utilize the bandwidth most efficiently. Ring on the other hand has the longest message depth of P-1 for communicators with P chips but is well-known to be bandwidth efficient for large messages. Thus, we expect the direct algorithm to perform best for small messages and the ring algorithm for large.

\begin{figure}[t]
	\centering
	\includegraphics[page=8, width=\columnwidth]{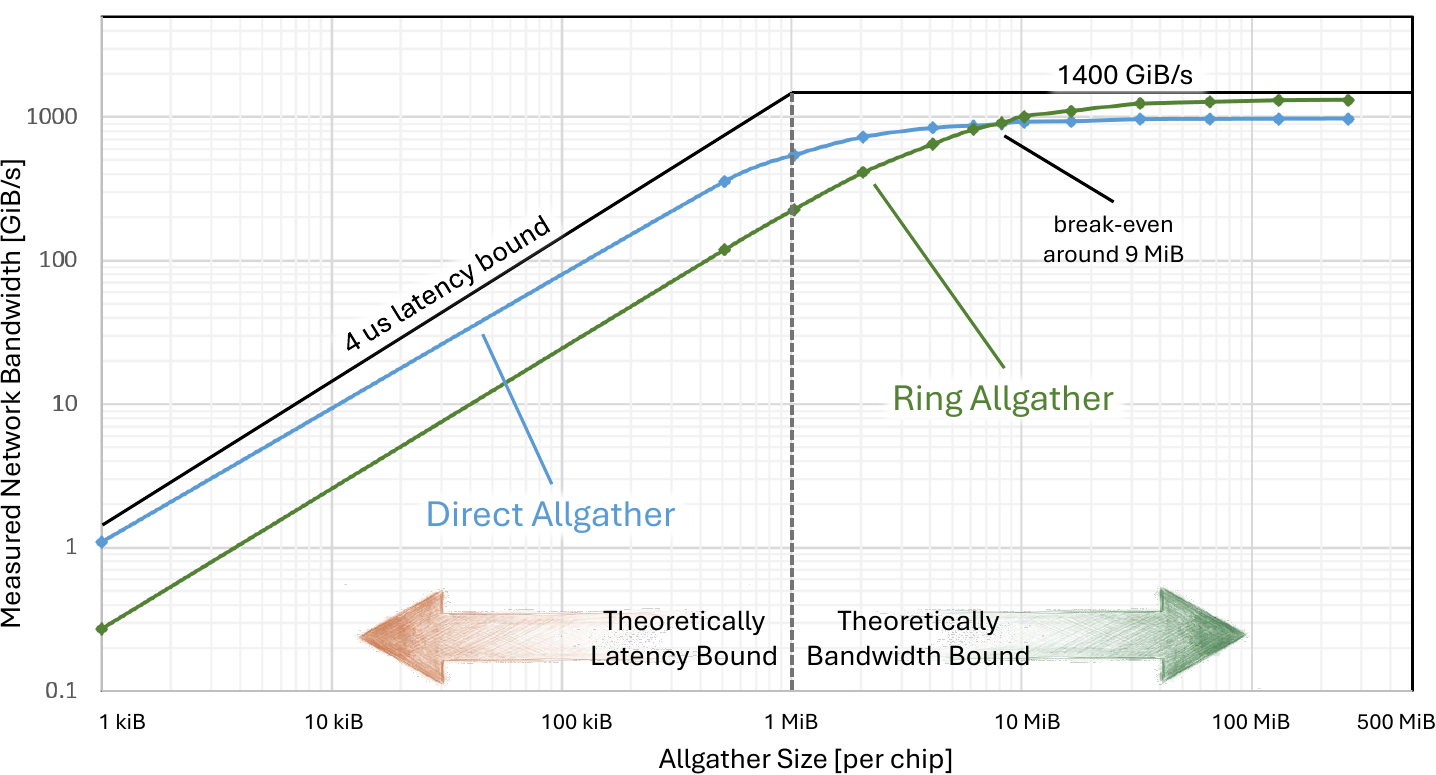}
	\caption{Allgather Benchmark Results on 8 Maia 200 Chips in Two Trays}
	\label{fig:allgather}
\end{figure}

Furthermore, we establish the \emph{speed of light (SoL)} bounds that either algorithm cannot exceed using a simple assessment based on latency and bandwidth. We estimate the maximum small-message latency of the network to be around 4us when traversing through the switched part. Fig.~\ref{fig:topo} shows the bandwidth configuration. While the tray design may seem asymmetric, i.e., north-south links can carry only 300 GB/s while east-west and diagonal can carry 350 GB/s, we note that the switched part of the network (400 GB/s) can flexibly be used to balance the bandwidth by using 50 GB/s bandwidth to strengthen the north-south links. This would lead to a fully balanced system of 350 GB/s for all four directions, leading to a total of 1.4 TB/s balanced network bandwidth. This switched design makes configurations more flexible and allows programmers to adjust mappings to different traffic requirements \cite{hxmesh}. 

We do not intend to provide a fully exhaustive benchmarking study but rather demonstrate the efficacy of SDLA for real-world workloads. Thus, we focus our analysis on one of the most important collective operations for LLM inference: Allgather is a central component in Fully Sharded Data Parallel LLM inference \cite{megatron, fsdp} and 3D parallelism using narrow datatypes \cite{zero}. In Allgather, all chips collect data from all other chips (e.g., weights or activations). Specifically, for a communicator of size $P$ and a communication volume of $N$ Bytes per process, each chip receives $R=N(P-1)$ Bytes. The speed of light for this operation would thus be $SoL = \max\left(4us, \frac{R}{1.4 \mathit{TiB/s}}\right)$. Fig.~\ref{fig:allgather} shows the speed of light as upper bound to the bandwidth (in black) and measured benchmark results on 8 chips for relevant data sizes in our production workloads. We achieve 78\% of the latency bound as well as 94\% of the bandwidth bound defined by the architectural limits, meeting or exceeding similar architectures. In practice, AI workloads run a multi-algorithm through a collective communication library similar to NCCL~\cite{nccl} or MCCL \cite{mccl} that chooses the fastest implementation based on the input parameters.

\subsubsection{Azure integration}

\begin{figure}[t]
	\centering
	\includegraphics[page=9, width=\columnwidth]{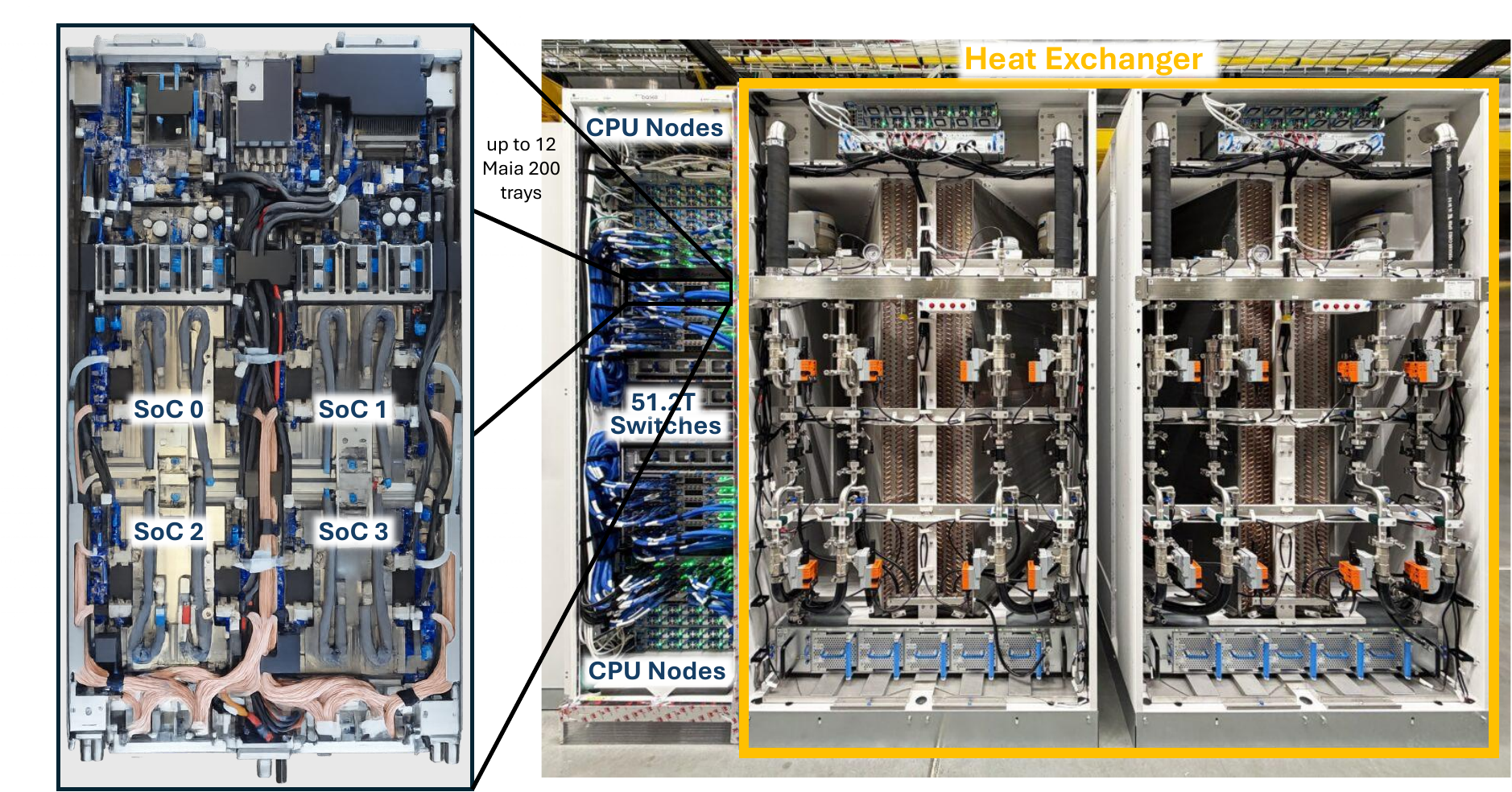}
	\caption{Maia 200 Tray (left) and Rack for Deployment in Air-Cooled Environments (right)}
	\label{fig:azure}
\end{figure}

Maia 200 is fully integrated into Microsoft’s Azure datacenters. The package is liquid cooled and can be connected to liquid cooling datacenter infrastructure. A tray is shown in the left part of Fig.~\ref{fig:azure}. Since liquid cooling is not standard in most datacenters yet, Maia 200 can be deployed in air-cooled datacenters by using an integrated heat exchanger as shown in the right part of Fig.~\ref{fig:azure}. Maia 200 uses standard Ethernet cabling and switches for all networking. This deep integration into existing Azure infrastructure contributes to the total TCO savings.

\section{End-to-end Generative Inference in a Real-World Setting}

In closing, we demonstrate a complete end-to-end example for inference running a public model, Qwen 2.5 7B~\cite{qwen} on a Maia 200 chip using all SDLA features described before. The model has 28 layers with an inner dimension of size \num{3584} and a feed-forward intermediate size of \num{18944}. It has 28 attention heads using grouped query attention with 4 KV heads. We write a matrix multiplication that consumes two input matrices of size N x K and K x M, respectively as shape N x K x M. Thus, the key and value projection matrix multiplications for a sequence with S entries are shaped S x \num{3584} x 512 and the query and output projection matrix multiplications are shaped S x \num{3584} x \num{3584}. The feed-forward layer’s up projection uses a SwiGLU activation function, which combines two matrix multiplications shaped S x 3584 x \num{18944} with a down projection of shape S x \num{18944} x \num{3584}. The final logits head projects to the vocabulary size with a matrix multiplication of shape S x \num{3584} x \num{151936}. Table~\ref{tab:qwen} shows the matrix multiplications and their shapes, equaling a total size of 14.14 GiB for all weights.

\begin{table}[t]
	\centering
			\caption{Qwen 2.5 7B Matrix Multiplication and Load Sizes}
\label{tab:qwen}\vspace{-6pt}
	\begin{tabular}{c | c | c | c }

						\rowcolor{tablegray}
				\textbf{Operation} & \textbf{Shape [N x K x M]}& \textbf{Count} & \textbf{Size [MiB]}  \\
		
				\hline

		k\_/v\_projection  & S x \num{3584} x 512 & 56 & 3.67  \\
		q\_/o\_projection  & S x \num{3584} x \num{3584} & 56 & 25.69   \\
		up\_/gate\_projection & S x \num{3584} x \num{18944} & 56 & 135.79  \\
		down\_projection & S x \num{18944} x \num{3584} & 28 & 135.79  \\
		logits & S x \num{3584} x \num{151936} & 1 & \num{1089.08}   \\

					\end{tabular}
	\vspace{-15pt}
\end{table}

For our analysis, we focus on the most important and most challenging memory-bound token generation phase. The arithmetic intensity in this case, where S=1, makes the workload fully memory bound and loading 112 small matrices of less than 26 MB and 84 medium-sized matrices of 136 MB and one large matrix of about 1 GB. The table also shows the expected load bandwidth according to our roofline plot in Fig.~\ref{fig:bf16}. We analyze inference for the case of a long generation of \num{16384} previous tokens to generate the \num{16385}th token autoregressively. The size of the KV cache in this case is 2 (K, V) x 28 (layers) x 4 (heads/layer) x 128 (dim per head) x 2 (Bytes / dim) x \num{16384} = 939.52 MiB. 
Our straight-forward implementation orchestrated with PyTorch calling standard kernels without additional fusing achieves 2434 tokens/s, which is more than 70\% of the estimated maximum performance. We also validated our implementation for correctness with respect to existing GPU inference solutions. 

This demonstrates that Maia 200’s implementation of the software-defined locally accessed (SDLA) dataflow architecture principle performs well not only close to peak for single matrix multiplication or collective operations but also for complex end-to-end practical inference workloads. This first feasibility study and implementation open the door for many additional optimizations to inline operators minimizing data movement. While we achieve good baseline performance, SDLA enables ninja programmers to explicitly manage every memory allocation and movement and orchestrate asynchronous control codes to achieve near-100\% utilization of all execution units even for complex workloads similar to what we demonstrated for matrix multiplications. 

\section{Related work}

Many new AI accelerators are entering the market to serve the quickly growing demand of compute cycles. In addition to many independent solutions~\cite{wse,cambricon}, several datacenter providers have their own designs: AWS’ Trainium and Inferentia~\cite{trainium}, Google’s TPUs~\cite{tpu1,tpu2}, and Meta’s MTIA~\cite{mtia}. Established GPU accelerators follow suit and specifically optimize for AI and LLM workloads. Our work follows and extends this lineage of existing accelerators towards software-defined dataflow using localized memories. 

One could argue that modern GPUs are moving in the same direction: NVIDIA's Tensor Memory Accelerator (TMA), introduced in Hopper GPUs~\cite{tma}, can be seen as comprising specialized memory and data movement engines to be hand-orchestrated by programmers. Yet, TMA interacts in non-trivial ways with the GPUs SIMT/warp system leading to a complete redesign of Hopper's warp-group MMA (WGMMA) to Unified (UMMA) in Blackwell, breaking backwards compatibility~\cite{nvidia_ptx_92}. 
In SDLA, we made explicit and localized memory programming a foundational principle that fits seamlessly from day one leading to improved energy efficiency.

Furthermore, one can argue that performance-aware AI programmers are moving towards more explicit, manually-scheduled approaches where they work around hidden architecture features. One of the prime examples is the series of four Flash Attention papers that carefully adjust the kernels to each new GPU architecture in a scientific paper each~\cite{fa1,fa2,fa3,fa4}. Such manually-scheduled ninja programming is becoming the new norm and frameworks like Triton~\cite{triton} or DaCe~\cite{dace} attempt to fill the resulting productivity gap. SDLA explicitly enables lowest-level control to simplify this transition.

\section{Discussion and Conclusions}

We discussed Software-Defined Locally Addressed Dataflow as a new architectural model to co-design and build efficient AI and HPC accelerator systems. SDLA focuses on data movement and its orchestration as leading design principles. We establish a taxonomy, similar to Flynn’s taxonomy that illustrates how SDLA combines two foundational principles: (1) \textbf{software-defined dataflow} that explicitly orchestrates the data path, allowing programmers to achieve more than 99\% utilization of tensor units by offloading the control and (2) \textbf{local data access} using optimized on-chip and system-wide memory placement that allows designers to build and co-design machines to specific dataflow requirements of the application. We also describe and demonstrate Maia 200, a real-world implementation of SDLA deployed in production in Microsoft’s fleet today. We show how Maia combines specialized data movement through distributed memories and networks and specialized control through explicit DMA, Synchronization, and Control units into a highly efficient architecture. 

\section*{Acknowledgments}

\textbf{No} text in this document was produced by an AI or LLM. Arrows in Figs. 4+5 were generated using a diffusion model.

\balance
\bibliographystyle{IEEEtranS}
\bibliography{maia_arXiv}

\end{document}